\documentclass[preprint]{iopart}

\usepackage{amsmath}
\usepackage{amsfonts}
\usepackage{amssymb}
\usepackage{graphicx}
\usepackage[T1]{fontenc}
\usepackage{url}
\renewcommand\d{\mathrm{d}}
\usepackage{tikz}
\usepackage[mode=buildnew,subpreambles=true]{standalone}
\begin{document}

\title{In Praise of Slowness}

%When submitting the manuscript for review, do not include the author's name or institution
\author{Roland Lehoucq}
\address{Université Paris-Saclay, Université Paris Cité, CEA, CNRS, AIM, 91191, Gif-sur-Yvette, France \\\url{roland.lehoucq@cea.fr}}
\author{Etienne Ligout}
\address{Astroparticule et Cosmologie, Université Paris Cité, 75013 Paris, France\\ \url{etienne.ligout@apc.in2p3.fr}} % optional
% ORCiD: 0000-0002-1618-2107

% If there were a second author at the same address, we would put another 
% \author{} statement here.  Don't combine multiple authors in a single
% \author statement.

% See the REVTeX documentation for more examples of author and affiliation lists.

\date{\today}

\begin{abstract}
 While speed is an ubiquitous concept in physics, its inverse — known as slowness — sometimes proves more relevant. We discuss some case studies within classical physics where such a notion is fruitful, before exploring how it can be realized within the relativistic and thermodynamical frameworks. This leads us to a generalized slowness, not specifically defined in relation with speed but as characterizing the evolution of any physical parameter. We advocate the pedagogical relevance of slowness, by analogy with other couples of quantities, inverse of each other, commonly used in physics.   

\end{abstract}

\maketitle % title page is now complete

%==================================================================
\section{Introduction} % Speed is not always relevant
\label{sec: Introduction}

\par Speed — the scalar quantity characterizing the rate at which an object covers a given distance — is so prevalent in physics that its legitimacy is rarely brought into question. Most often, one indeed regards speed as \emph{the} natural concept to describe the motion of a physical system, defined as the ratio $v=\Delta r/ \Delta t$, \emph{i.e.} the ratio of the distance $\Delta r$ traveled over the time $\Delta t$ it took the system to cover that distance. Here, we actually are talking about the average speed of the system, but one could also consider the instantaneous speed. Whichever realization of the notion of speed is used, the core idea remains the same: a ratio of distance over time. But the puzzle then comes from the apparent arbitrariness of the order of this ratio. Why should we consider distance over time rather than time over distance? This comes down to asking why we should use speed and not the reciprocal scalar quantity, \emph{slowness}. After all, to say that a system at rest has vanishing speed or infinite slowness is but a terminological tweak, just like saying that a system covering a large distance in a small amount of time has a large speed or a small slowness (and \emph{vice versa}).

\par A first obvious explanation of the prevalence of speed pertains to the historical contingency of the development of science: speed came to be used instead of slowness, and this is why it seems so natural compared to the latter quantity. This answer is not satisfactory though, because it does not explain the gradual uptake of speed in the physical sciences since the $17^{\rm th}$ century. %\footnote{[\textbf{EL : supprimer cette note ? RL : OUI}] Taking the ratio of two quantities of different dimensions was not always so natural as it seems nowadays. In the $17^{\rm th}$ century, only dimensionless ratios were used; strictly speaking, the notion of speed (understood as a quantity with a new physical dimension, \emph{e.g.} expressed in $\mathrm{m}/\mathrm{s}$) did not exist back then and one instead spoke of the relation between the ratios of the distances covered and the time it took to cover them. In his \emph{Discorsi}, Galileo for instance characterized Aristotle's free fall law as follows: “bodies of different weight [...] move in one and the same medium with different speeds which stand to one another in the same ratio as the weights; so that, for example, a body which is ten times as heavy as another will move ten times as rapidly as the other.”; M. Schrenk, “Galileo versus Aristotle on Free Falling Bodies”, Log. Anal. Hist. Philos. \textbf{7}(1), 81--89 (2004).}
This leads us to a second potential explanation: speed is preferred over slowness simply because it is more convenient. At first glance, this view seems justified, \emph{e.g.} in mechanics where kinetic energy is more naturally expressed using $v$ than its inverse. More fundamentally, it is acceleration -- the time derivative of speed -- that matters for dynamics: Newton's second law of motion links force and acceleration. At that level, there is a fundamental asymmetry between space and time: the effect of a force is a second derivative of space in time, which is certainly distinct from the second derivative of time over space. That being said, in the context of kinematics, discussing an alternative way to look at motion  remains fruitful, in that it allows us to have a fresh look at some infamous problems. In many such instances, the study of a moving system implies a fixed time $\Delta t$ (of the experiment undertaken or the phenomenon observed) and a yet to be determined physical distance $\Delta r$ (the characteristic length of the motion). When this is so, it is more convenient to consider the speed of the system at the various intermediary times than its slowness at the same times. Indeed, if over $\Delta t$ the system (moving in one dimension, for the sake of simplicity) sees its speed altered $N$ times, so as to make it move to speed $v_i$ in the time interval $\Delta t_i$ (for $i=1,\dots, N$), then the time-averaged speed of the particle over the entire path is:
\begin{equation}
\label{equation average speed first example intro}
    \bar{v} = \frac{1}{\Delta t} \sum_{i\,=\,1}^{N} v_i\, \Delta t_i.
\end{equation}
Clearly, using $s_i=1/v_i$ and the resulting average slowness $\overline{s}$ would lead to a rather awkward sum here, which explains why speed is favored. But the point is that in some situations the converse happens: \emph{speed is sometimes less convenient than slowness}. To understand why, note that in some physical situations the natural slicing of the path of the system is not temporal but spatial: this typically happens when we know where the system will end up, so that the distance covered is fixed, but we do not know when it will reach that endpoint. By symmetry with the previous case, $\Delta r$ will be sliced into $N$ subdistances $\Delta r_i$ (still with $i=1,\dots, N$) each associated to a specific speed $v_i$. The average speed is then no longer a weighted arithmetic mean like in Eq.~\eqref{equation average speed first example intro}, but a weighted harmonic mean of the form 
\begin{equation}
\label{equation average speed second example intro}
    \bar{v} = \Delta r \left( \sum_{i\,=\,1}^{N} \frac{\Delta r_i}{v_i} \right)^{-1}.
\end{equation}
The occasional awkwardness of speed mentioned previously is now blatant. But if $\bar{v}$ proves inconvenient, we can consider the mean slowness $\overline{s}$ instead, which gives the much nicer weighted arithmetic mean: 
\begin{equation}
\label{average slowness second example intro}
      \bar{s} = \frac{1}{\Delta r} \sum_{i\,=\,1}^{N} s_i\, \Delta r_i.
\end{equation}
Quite interestingly, the relation Eq.~\eqref{average slowness second example intro} — which the reader might deem to be a mere curiosity at the moment — is used in many fields outside of the physical sciences. Long-distance runners record “minutes per kilometer” during training or competitions; in this way, they need only add their slowness (“pace”, in their vocabulary) on each spatial segment of the path (\emph{e.g.} a kilometer or a leap) in order to obtain their global performance, which is far simpler than adding the reciprocal of their speed along each segment according to Eq.~\eqref{equation average speed second example intro}. Here, one might also draw an analogy with electrical circuits: just as for some circuits configurations, the additive quantity turns out to be the conductance rather than the resistance, for some kinematic settings the additive quantity turns out to be the slowness rather than the speed. In the same spirit, in a car race one always announces the time per lap of the drivers, not their speed.  The crux of the matter is that in many (if not most) daily applications, the distance to be covered is fixed and the key unknown factor is the time to cover said distance. This point is apparent in travel and navigation: commuters and travelers are not so interested in the distance they cover than in the time the trip takes. During the colonial era in Mexico ($16^{\rm th}-18^{\rm th}$ centuries), people used
the league, a unit of distance worth 4.19 km, to refer to the distance covered by foot in one hour on a flat and even ground; however, this league was sometimes said to be ``short'' or ``long'' depending on the topography on the route taken, which shortened or lengthened the journey time \cite{Garza2012}. Moreover, global navigation satellite systems (GNSS) do not calculate the trajectory of least distance, but the trajectory of least time. Lastly, in aviation, one commonly uses the time  per nautical mile, that is the inverse of the speed (expressed in knots) — to calculate the aircraft's flight time and maximum drift induced by the wind. If then slowness proves more convenient than speed in such daily applications, this raises the question of the overall relevance of this notion in physics.\footnote{In some contexts, a quantity and its inverse share the same relevance. For instance, in the US, automotive fuel consumption is expressed in miles per gallon (distance per unit volume) while Europe uses the inverse quantity (unit volume per distance) expressed in liters per 100 kilometers.} This is the issue we want to tackle here, so as to demonstrate that slowness can be a valuable tool in some physical problems, and in such cases more useful than the notion of speed.

%\par The paper is organized as follows. In section 2, we detail how the notion of slowness naturally appears in several optimization problems of classical physics, ranging from optics, through seismology, to sailing. In section 3, we detail how slowness can be framed in relativity and thermodynamics, and flesh out from this analysis a generalization of the initial notion (related to speed) pertaining to the time evolution of any physical parameter.

\section{Slowness in optimization problems}

\par In classical physics already, one encounters many situations where speed is not the most relevant parameter. This typically happens in optimization problems, where one is more interested in the duration of the phenomenon at hand than in the distance on which this phenomenon happens. In this kind of problem, both the initial and final endpoints of the path followed by the system (be it a path in physical space or a more abstract one) are fixed, so that the goal is not to know where the system will end, but only when it will end at the (fixed) final point. Regardless of the specifications of the physics involved, in this context the goal is always to find the optimal path. As we illustrate now, solving this kind of problem naturally leads to slowness.   

\subsection{Optics} \label{subsection optics}
Perhaps the simplest example of a situation where one is not so interested in the distance between two points is in geometrical optics, where refraction shows that light rays do not always follow a straight line. This elementary observation is formalized by Fermat's principle and more commonly referred to as the principle of least time: the path taken by a ray of light joining two points $A$ and $B$ is the one that can be traveled in the least time. If  $\Delta t_{AB}$ is the duration of this optical path, 
\begin{equation}
\label{duration between A and B optical path (first equation)}
     \Delta t_{AB} = \int_{t_A}^{t_B} \mathrm{d} t =  \int_{\ell_A}^{\ell_B} \frac{\mathrm{d} \ell }{v}\,
\end{equation}
 where $\d \ell$ an infinitesimal displacement along the ray and $v = \d \ell/\d t$ the speed of light in the medium considered. One usually introduces the refractive index of the medium (which can be a function of position) $n = c/v$, $c$ being the speed of light in the vacuum. Fermat's principle states that the path actually followed by the ray is the one that minimizes $\Delta t_{AB}$, in the sense that any variation to said path would lead to a longer duration of travel. But note that Eq.~\eqref{duration between A and B optical path (first equation)} explicitly incorporates the slowness of light $1/v$. Therefore, the principle of least time can also be interpreted as a principle of \emph{least slowness}: the light follows the path where it will be the least slowed down. This unorthodox way of looking at Fermat's principle — and the associated terminology of “light slowness”\footnote{This term is mentioned in one instance in O. Darrigol, \emph{A History of Optics}, Oxford University Press, Oxford (2012), p. 269. Thus, using the expression “light slowness” is not only infrequent in the modern literature, but also in the first studies on optics.} — is but a slight shift on the standard perspective, which favors the concept of the (infinitesimal) optical length  $n\, \d \ell$. 
 
\par One may object to what we have just said that Eq.~\eqref{duration between A and B optical path (first equation)} could simply be understood as depending on the parameter $v$, it appearing as an inverse being of no significance. To which extent focusing on $1/v$ instead of $v$ is insightful ? For it is obvious that \emph{any} physical parameter $A$ of a theory can be understood in terms of its reciprocal $1/A$ without altering anything in the mathematical framework: we are always free to choose the variables used. However, this does not imply that the theory will be \emph{as simple} when expressed in terms of $1/A$ or in terms of $A$. Applied to geometrical optics, this asymmetry in simplicity provides a good reason to regard $1/v$ — quantified by $n$ — as the most natural variable. The Euler-Lagrange equations of Eq.~\eqref{duration between A and B optical path (first equation)} lead to the eikonal equations
\begin{equation}
\label{eikonal equation optics}
    \frac{\mathrm{d}}{\mathrm{d} \tau} \left(n \,   \frac{\mathrm{d} x^i}{\mathrm{d} \tau} \right) = \frac{\partial n}{ \partial x^i},
\end{equation}
where $x^i=(x,y,z)$ are cartesian spatial coordinates and $\tau$ parametrizes the path. Evidently, writing \eqref{eikonal equation optics} in terms of $v \propto 1/n$ is more intricate. One does not \emph{have} to frame his view of optics primarily in terms of slowness  instead of speed, but it certainly provides a simpler dynamics. The crux of the matter is the following: in physics, the relevance of a quantity is solely justified by its convenience. And, aptly, the slowness, \emph{via} $n$, appears with the same convenience in various extremal principles at the foundations of optics: beyond that of Fermat, one can think of the Huygens' and Maupertuis' principles \cite{Kimball1998}.

\subsection{Seismology}

\par Slowness can not only be used to understand the propagation of light, but also that of sound. A field of study particularly interested in sound waves is seismology, where the concept of slowness is commonly used and preferred to that of speed. Seismologists typically deal with waves traveling in the strata of media that constitute the layers of the Earth, which are assumed to be horizontal stacks (indeed, such waves reach depths of the order of $10^1-10^2\,\mathrm{km}$, which is small compared to the radius of curvature of the Earth). Hence, they will resort to the Snell-Descartes' law: given two media with acoustical indices $n$ and $n'$, the angles of incidence and refraction, $\theta$ and $\theta'$, are related by $n\,\sin (\theta) =  n'\, \sin (\theta')$, which can be rewritten as
\begin{equation}
\label{snell law first form}
     \frac{1}{v} \,\mathrm{sin}(\theta) =  \frac{1}{v'} \, \mathrm{sin}(\theta'),
\end{equation}
where $v$ and $v'$ respectively are the speed of sound in the medium of indices $n$ and $n'$. $\theta'$ is also the angle of incidence at the next interface of the strata (for our purpose, the next boundary between layers of the Earth); thus,  Eq.~\eqref{snell law first form} says that the horizontal component of the inverse of speed, $p=(1/v) \,\mathrm{sin}(\theta)$, stays constant throughout the propagation of the wave in various media. This conserved quantity is called the \emph{horizontal slowness} \cite{Kennett1981}. As we dive through the layers of the Earth, the  average density increases and so it is quite natural to assume that the speed of sound $v$ does too. $1/v$ will then decrease as we go deeper, which can only be compensated in the law $p = \mathrm{constant}$ if $\theta$ gets closer to $\pi/2$. For a wave sent from the surface, this fact implies — provided a sufficient decrease in
$1/v$ — the existence of a turning point, at which the wave will be reflected back towards the surface (in optics, the same mechanism underlies the emergence of mirages or propagation of light in optical fiber).

\par Consider then a seismologist $S$, positioned on some place at the surface of the Earth, studying a wave sent into the ground by a colleague $S'$ in another laboratory. They communicate during the experiment, so that $S$ knows where the wave was sent from and at which time. $S'$ diligently calibrated the angle at which the wave was sent, so $S$ also knows its horizontal slowness $p$. In summary, after receiving the signal, $S$ is in possession of $3$ parameters: the travel time $T$ of the wave, the horizontal distance $X$ it has covered (\emph{i.e.} the distance between $S$ and $S'$) and its horizontal slowness $p$. Let $v(z)$ be the speed of sound at depth $z$ from the surface, which is required to be an increasing function.  $S$ aims at deducing from $T$, $X$, $p$ and $v(z)$ the depth of the turning point $z_{\rm tp}$. In fact, the natural variable in this problem is not $v(z)$, but the slowness of sound at depth $z$, $s(z)=1/v(z)$. To find $z_{\rm tp}$, one can indeed start from the relation \cite{Kimball1998}
\begin{equation}
\label{expression X (p, z)}
    X(p,z_{\rm tp})= 2p \int_0^{z_{\rm tp}} \left(s^2(z)-p^2\right)^{-1/2}\d z, 
\end{equation}
where the integrand is the reciprocal of the vertical slowness (which is \emph{not} the vertical speed).  To find an analogous expression for $T$, we can use the fact that the horizontal slowness is constant throughout the entire travel of the wave,  leading to the relation $p=\d T/ \d X$. Then, carefully distinguishing the functional dependence of $\mathrm{d}T$ on both $p$ and $z$, one finds
\begin{equation}
\label{expression T (p, z)}
    T(p,z_{\rm tp})= 2 \int_0^{z_{\rm tp}}  s^2(z)\left(s^2(z)-p^2\right)^{-1/2}\d z. 
\end{equation}
Evaluating Eqs.~\eqref{expression X (p, z)} and \eqref{expression T (p, z)} with the appropriate value of $p$ yields $z_{\rm tp}$.\\
\par Here,  slowness, by means of the parameter $p$ and the function $s(z)$, prevails over speed. Yet, Eqs.~\eqref{expression X (p, z)} and \eqref{expression T (p, z)} are not specific to seismology, but would also hold in optics, \emph{e.g.} in the analysis of mirages, which is concerned with the propagation of a ray shined on a strata of media of increasing refractive indices (upwards from the ground) towards a location at the same height as the starting point. With that in mind, how can we explain that slowness is a common term in seismology, but not in optics (in the conventional perspective of the latter field, beyond the points raised in the previous subsection) ? A first aspect is the set-up of seismology, which, considered from the broader view of the theory of wave propagation, is rather specific. This field primarily focuses on the Earth, which amounts to a set of concentric strata whose total extension (\emph{i.e.} the radius of the planet $R_{\rm T}$) is fixed. Therefore, in describing the propagation of a sound wave, we will resort to a spatial slicing of $R_{\rm T}$ into smaller distances $r_i$ (for $i=1,\dots, N$). The point is not so much to determine where the wave will end up (since we know it will travel through the Earth and come back at the surface eventually) but rather to determine when it will reach the endpoint. In the introduction, we showed that in this kind of setting slowness is more convenient, because it leads to more manageable expressions (see Eq.~\eqref{average slowness second example intro}). Equations \eqref{expression X (p, z)} and \eqref{expression T (p, z)} illustrate this point clearly: $v(z)$ would be more awkward to use therein than $s(z)$. Moreover, slowness seems natural here for it yields to a conserved quantity, namely the horizontal slowness $p$. It makes sense to frame the physics at stake in terms of the relation between $p$ (the conserved quantity) and $s(z)$ (the varying quantity), just like we usually frame mechanics in terms of the relation between energy and various varying quantities (\emph{e.g.} the position and speed of a moving particle).

%\begin{figure}[h!]
%    \centering
%\includegraphics[scale=1.3]{Figures/sismology.pdf}
%    \caption{A seismologist $S$ sends a wave thought the layers of the Earth (sketched as blue lines being refracted successively through the media), which is reflected at the turning point (at depth $z_{\rm tp}$) and is received by seismologist $S'$. }
%    \label{fig:enter-label}
%\end{figure}

\section{From relativity to thermodynamics: generalizing slowness}
\par In the previous section, slowness was shown to be a fruitful notion in the rather specific context of optimization problems in non-relativistic mechanics. As we will now see, this concept can be motivated in the broader context of pivotal branches of classical physics: special and general relativity, first, and also thermodynamics. This discussion leads us to flesh out the notion of generalized slowness.
\label{sec:Generalisation}

\subsection{Relativistic motion}

\par One often hears that in special relativity, space and time have the same status. This is not so accurate, due to a fundamental asymmetry: whereas a given physical system can stay at the same place, it can never stay at the same time. This fact is reflected on (two-dimensional) Minkowski diagrams \cite{Minkowski1908} insofar as they are drawn in the $(r, ct)$ plane, not the $(ct, r)$ plane: time is steadily flowing upwards, never in the reverse direction. This graphical representation of spacetime then does, in itself, suggest the relevance of slowness to describe the motion of bodies. Indeed, the slope of a wordline is (up to a factor $c$) \emph{not} the speed $v=\mathrm{d}r/\mathrm{d}t$, but the slowness $s=\mathrm{d}t/\mathrm{d}r$. Thus, the slope of the wordline of a massive body can never exceed $1/c$, the slowness of light in the vacuum. In the Newtonian limit, this maximal slowness is sent to 0 to reduce the Lorentz group to the Galilean one. The symmetrical situation, an infinitely slow light ($c \rightarrow 0$) — as weird as it may appear at first glance — corresponds to the so-called Carroll group \cite{JMLL1965}. This is the limit explored with humor by George Gamow in his famous 1940 book \emph{Mr Tompkins in Wonderland}, where the hero, C.~G.~H. Tompkins, enters a dream world in which the speed of light is a mere 10 mph. 
%\footnote{A ``Carrollian physics'' is now developing, with applications in various domains of forefront theoretical physics such as quantum gravitation, supersymmetry or string theory. See the recent second Carrol workshop (Mons, September 2022).}
%\url{<https://web.umons.ac.be/pucg/event/carroll-workshop-2>}

%It should also be noted that in special relativity, there are three distinct operational procedures to define speed:\cite{JMLL1980}
%\begin{itemize}
%    \item the usual Galilean speed $v$, defined as the ratio of distance to duration measured in the same reference frame;
%    \item the ``celerity'' $w$, allowing to express the momentum $p$ of a particle of invariant mass $m$ in a Galilean style, $p = m w$;
%    \item The ``rapidity'' $\alpha$, which is the additive parameter of the motion, obeying a linear combination law $\alpha_{12} = \alpha_{1} + \alpha_{2}$ identical to that of the Galilean speed in Newtonian mechanics.
%\end{itemize}
%Adding slowness to this list of mathematically equivalent quantities may help to clarify conceptual subtleties.\\

%In the famous Langevin twin thought experiment, Alice (who stay on Earth) send periodic messages, with proper period $T_A$ to Bob who is traveling way with speed $\beta c$. He received them with proper period $T_B$. The ratio $T_B/T_A$ between these durations is equal to the Dopler factor $\sqrt{(1+\beta)/(1-\beta)}$.

\par
\begin{figure}[h!]
\centering
    \includegraphics[width=0.8\linewidth]{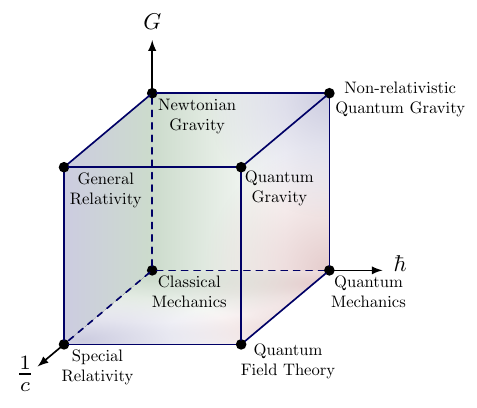}
      \caption{Cube of physical theories.}
    \label{figure cube}
\end{figure}

 The foregoing only points to the possibility of slowness in special relativity. While one could indeed interpret the slope on a Minkowski diagram or even the Lorentz factor $\gamma$ as realizations of the concept of slowness, such tweaks are not very illuminating \emph{per se}. A more convincing route to slowness can be identified once we depart from the idea of the slowness of a physical system (associated to a specific motion) to focus on a fundamental slowness. In deriving the Lorentz transformations from general physical principles, at some point we find the transformation laws~\cite{JMLL1976} 
\begin{equation}
\label{Lorentz transformations form}
    r' = \left(1-\alpha_{0} V^2 \right)^{-1/2} \left(r-V t \right), t' = \left(1-\alpha_{0} V^2 \right)^{-1/2} \left(t-\alpha_{0} V r  \right),
\end{equation}
where $V$ is the speed of the $(r',ct')$ frame relative to the $(r,ct)$ frame and  $\alpha_{0} >0$ is a yet to be determined constant. We know of course that $\alpha_{0}=1/c^{2}$, and this is why Eq.~\eqref{Lorentz transformations form} is always expressed in terms of $c$. Still,  this rewriting of $\alpha_{0}$ is not as legitimate as it might seem at first. To understand why, it will prove convenient to use a graphical representation called the \emph{cube of theories}. Also known as the $c\,G\hbar$ cube, the cube of theories~\cite{Okun1991} (drawn on Fig. \ref{figure cube}) 
provides an elegant way to represent all regimes of physics, from classical mechanics to quantum gravity. In a three-dimensional space, we locate classical mechanics (without gravitation) at the origin and place all regimes stemming from the constants $c$, $\hbar$ and $G$ (and their various combinations) at the vertices of an unit cube. On each axis, the constant is turned off at the location $0$ (origin), and turned on at the location $1$ (a vertex of the cube).

\par  This cube highlights the fact that \emph{the fundamental constant of special relativity}, analogous to $\hbar$ for quantum mechanics and $G$ for gravity, \emph{is $1/c$ rather than $c$}. To be sure, the intensity of an interaction is more conveniently characterized by a bounded constant, if we are to turn it off and on. Quantum and gravitational effects respectively become negligible when $\hbar \rightarrow 0$ and $G \rightarrow 0$; yet, to turn off the special relativistic effects, one usually sets $c \rightarrow \infty$. To amend this discrepancy, we should view $s_0=1/c$, which can be interpreted as the slowness of light in the vacuum, as the appropriate relativistic constant. More fundamentally however, $s_0$ is the \emph{lowest physical slowness}; this perspective goes hand in hand with a systematic interpretation of the fundamental constants of physics as constraints (lower limits, in fact) on observable phenomena \cite{Lajze2018}. Beyond the historical context of derivation of the Lorentz transformation laws, there is no  reason to write $\alpha_0= s_0^2$ (a fundamental constant of physics, a geometrical parameter of Minkowski spacetime in a sense) as an inverse. The fact that $s_{0}$ is the 
proper relativistic constant also transpires in general relativity, for corrections to Newtonian gravity are quantified by (powers of) $s_0$.  An astrophysical body of mass $M$ and radius $R$ can for instance be described classically as long as $R \gg 2GM s_0^2$. The description of the inspiral phase of a binary merger relies on the so-called post-Newtonian approximation scheme, which expands the metric tensor in powers of $s_0$, to find approximate solutions to the Einstein equations \cite{Poisson2014}.

\subsection{Time evolution and generalized slowness}
\label{subsection thermodynamics}
\par Thermodynamics relies on the fundamental assumption that the system considered is at equilibrium at the beginning and at the end of the process studied. The entire point is then to track down the evolution of the thermodynamical parameters characterizing the system (volume, pressure, temperature, charge, etc.) during the process.
But, crucially, such thermodynamical parameters are defined only at equilibrium.   Thus, the state of the system during the process can only be described thermodynamically if said process is quasi-static (\emph{i.e.} reducible to a succession of intermediary equilibrium states). In practice, a process is modeled as quasi-static if it is “slow enough”. What underlies this rather vague statement is a comparison between the characteristic time $\tau_{\rm p}$ associated to the (macroscopic) process and the characteristic time $\tau_{\rm r}$ associated to the relaxation of the system (the time it takes to go back to equilibrium after being suddenly perturbed). The quasi-static regime proves adequate whenever $\tau_{\rm r} \ll \tau_{\rm p}$. from a mathematical standpoint, an idealization is involved here: quasi-staticity does not mean “slow enough”, but “infinitely slow”. In real-world applications, the system is never strictly speaking at equilibrium because  $\tau_{\rm r} \neq 0$, but can be considered so as long as $\tau_{\rm p} \gg \tau_{\rm r}$. Even if infinite slowness is an idealization of the intricate relation between the system and its environment (as quantified by the relative order of magnitude between $\tau_{\rm p}$ and $\tau_{\rm r}$), the slowness of processes 
remains an important notion in thermodynamics. This point is reflected in the terminology, insofar as processes are never said to be ``\emph{not too} fast” but rather to be ``slow \emph{enough}”. Comparing the characteristic times associated to the various physical phenomena responsible for the evolution of a system comes down to comparing the slowness of such phenomena. In this sense, the longer the characteristic time, the greater the slowness. Most equilibrium thermodynamics can be understood in this manner. As a simple illustration, consider the rheological issue of the characterization of the fluidity of materials under given conditions. A convenient quantity in that context is the dimensionless Deborah number\cite{Reiner1964}, which quantifies both the propensity of a solid-like material to flow on a large time scale and the propensity of a fluid-like material to have properties of solids when it is deformed abruptly. In both cases, the Deborah number $\mathrm{De} =\tau_{\mathrm {r}}/\tau_{\mathrm {p}}$ is the ratio of two characteristic times, $\tau_{\mathrm {r}}$ (associated to the relaxation of the system after the application of a stress) and $\tau_{\mathrm {p}}$ (associated to the experiment undertaken to probe the response of the material to said stress). The goal here is to specify the relative order of magnitude between the duration of both phenomena, which comes down to a comparison of their slowness.  \\

\par From the foregoing, the relevance of slowness in thermodynamics appears quite broad, because one can consider the slowness of \emph{any} phenomenon (by means of its characteristic time); this notion can be devised for any kind of physical quantity, not just for distance as was done in most of the previous development. Such a \emph{generalized slowness} is motivated as follows. Let $Q$ be a given physical quantity characterizing the thermodynamical system considered. If the system undergoes an evolution (\emph{e.g.} by interacting with its environment), $Q$ is going to evolve more or less rapidly, depending on the characteristic time $\tau_{Q}$ associated to the phenomenon underlying the evolution of the system. Therefore, at least locally in time, 
\begin{equation}
\label{slowness link with characteristic time}
  \bigg\rvert  Q\,\frac{\mathrm{d}t}{\mathrm{d}Q}  \bigg\rvert = \tau_{Q},
\end{equation}
which is a consequence of the definition of $\tau_{Q}$ as the inverse of the logarithmic derivative of $Q$. Equation \eqref{slowness link with characteristic time} makes it clear that $\mathrm{d}Q / \mathrm{d}t$ is an analog of speed for the quantity $Q$; accordingly, $\mathrm{d}t / \mathrm{d}Q$ is analogue to the inverse of a speed, that is the sought-after generalized slowness. If $Q$ is for example the concentration of a specific chemical component during a reaction, then $\mathrm{d}t / \mathrm{d}Q$ will be the slowness of the concentration (evolution). Generalized slowness is especially useful in the context of energy extraction and consumption (in spite of not being called so therein). 
According to Hecht \cite{Hecht2019}, energy is a measure of the capacity of matter interacting with matter to perform physical change. But for industrial purposes, only steady energy flows are interesting: energy has to be injected into the system at a rate greater than that associated with the intrinsic losses of the system (\textit{e.g.} induced by conductive, convective or radiative transport). This means that slowness of energy, $\mathrm{d}t/\mathrm{d}E$, must be low enough; in more familiar terms,  power must be sufficiently high.  When considering the consumption of a finite resource like oil, the characteristic depletion time carries the same idea. If $R$ is the estimated remaining amount of the resource (usually expressed in tons) and $P$ its rate of extraction (usually expressed in tons/year), the characteristic depletion time $T$ is the reserves-to-production ratio $R/P$. While $T$ does \emph{not} represent the number of years before which the resource will be entirely depleted (because one must take into account future reserve growth and potential variations in the means of extraction), it still gives a valuable estimate: we are more interested in the time remaining before the resource eventually disappears than in the rate at which it is currently consumed.\\

\par Slowness is not just relevant to describe the relation between space and time (as the inverse of a speed), but is also fruitful to specify the relation between \emph{any quantity} and time. The underlying motivation for this concept --- here realized in the form of generalized slowness --- remains the same as before: fundamentally, what we are interested in is the characteristic duration of the evolution of the quantity $Q$ involved, not in its variations \emph{per se}. Basically, the flow rate $\mathrm{d}Q / \mathrm{d}t$ (quantity per unit time) is not easy to grasp compared with the slowness $\mathrm{d}t / \mathrm{d}Q$ (time per unit quantity), which in the end explains the legitimacy of the latter notion. 
\section{Conclusion}

\par We have seen that the notion of slowness, while quite obscure, proves to be relevant in various physical situations (and sometimes even more so than speed). Intuitively defined as the inverse of speed, this notion actually takes several forms depending on its context of use. We fleshed out a general definition of slowness relating to the evolution of any physical quantity, and not just to the evolution of the position of the system. Of course, the perspective adopted in this article is a thought exercise, with its limits. Speed certainly prevails over slowness at the dynamical level (as explained in the introduction). Still, we are lead to wonder why we do not teach students about speed \emph{and} slowness, although they often come to learn to manipulate similar couples of \emph{dual notions} — inverse of each other. If we come to learn about compressibility \emph{and} bulk modulus, resistance \emph{and} conductance, viscosity \emph{and} fluidity, wavelength \emph{and} wavenumber, it's because each element of such couples is the appropriate tool to solve specific physical problems. For this reason, we believe that the acknowledgment of slowness  is part of the — well, slow — process of improving our understanding and teaching of physics. As Voltaire said \cite{Voltaire1773}: ``[l]e monde avec lenteur marche vers la sagesse.'' (``The world  slowly walks towards wisdom.'')\\

\end{document}